\documentclass[10pt,twocolumn,letterpaper]{article}

\usepackage{iccv}              % To produce the CAMERA-READY version
\usepackage{booktabs}
\usepackage{xcolor}
\usepackage{colortbl}
\usepackage{multirow}
\usepackage{pifont}
\usepackage{tcolorbox}
\usepackage{listings}
\usepackage[accsupp]{axessibility}

\usepackage[misc]{ifsym}

\definecolor{iccvblue}{rgb}{0.21,0.49,0.74}
\usepackage[pagebackref,breaklinks,colorlinks,allcolors=iccvblue]{hyperref}

\def\paperID{9320} % *** Enter the Paper ID here
\def\confName{ICCV}
\def\confYear{2025}

\title{DanceEditor: Towards Iterative Editable Music-driven Dance Generation with Open-Vocabulary Descriptions}
\author{Hengyuan Zhang$^{1,\ast}$, Zhe Li$^{1, \ast}$, Xingqun Qi$^{2, \textrm{\Letter}}$, Mengze Li$^{2}$, Muyi Sun$^{3}$, Man Zhang$^{3}$, Sirui Han$^{2, \textrm{\Letter}}$\\
$^{1}$ Peking University,
$^{2}$ The Hong Kong University of Science and Technology \\ 
$^{3}$ Beijing University of Posts and Telecommunications\\
}

\begin{document}

\twocolumn[{%
    \renewcommand\twocolumn[1][]{#1}  % 避免标题被影响
    \maketitle  % 生成标题
    \vspace{-7mm}
    \begin{center}
        \centering
        \includegraphics[width=0.9\textwidth]{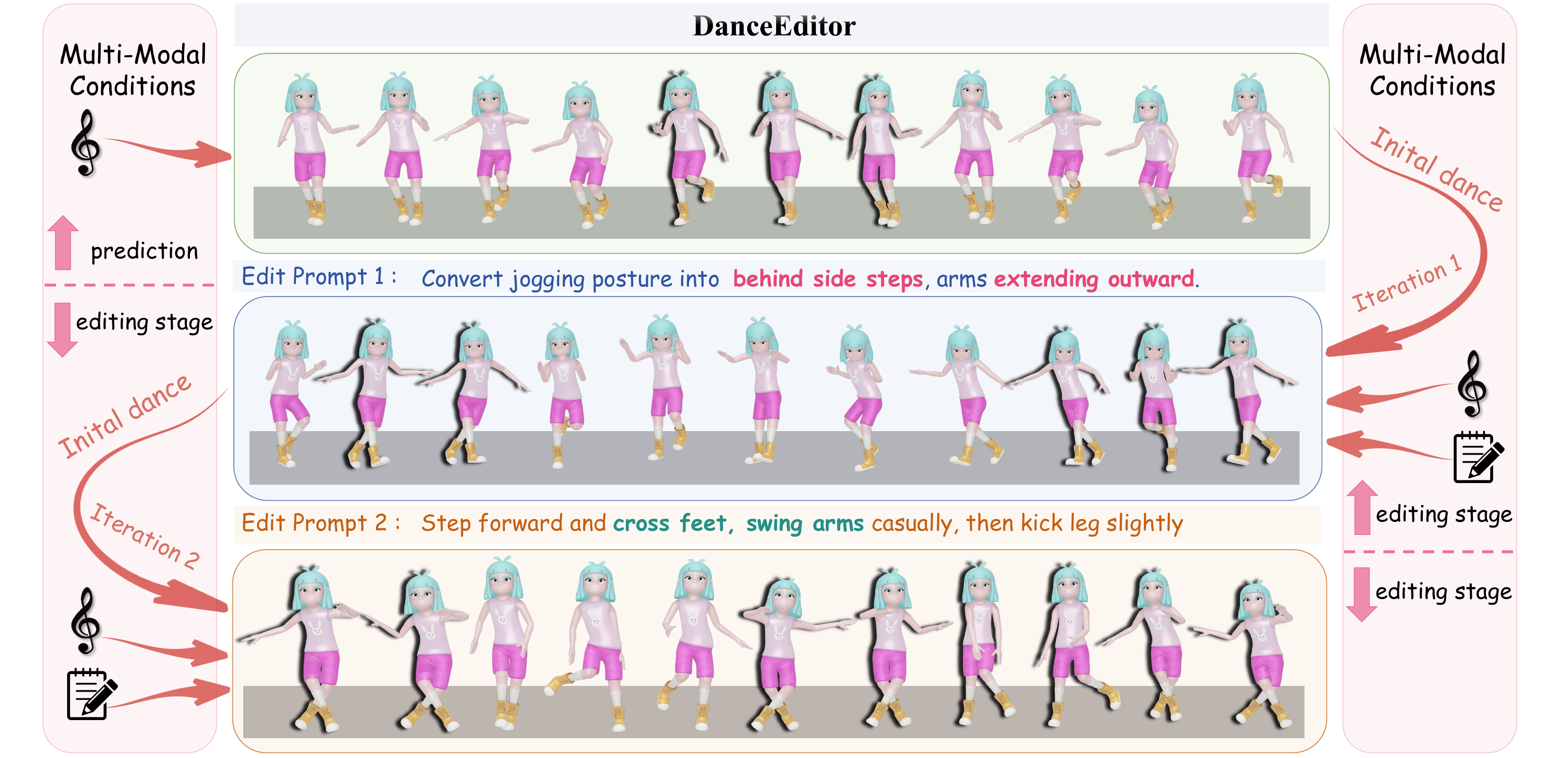}
        \captionof{figure}{ Our \textbf{DanceEditor} framework, pre-trained on a large-scale dataset, enables iterative and editable dance generation that is coherently aligned with the provided music signals. The highlighted texts and avatar shadow effects here specifically indicate edits related to body movements.}
        \label{fig:fig_teaser} 
    \end{center}
    % \vspace{-1mm}  % 适当增加 teaser 和正文的间距
}]

\renewcommand{\thefootnote}{\fnsymbol{footnote}}
\footnotetext{$\ast$ These authors contribute equally.
$\quad \textrm{\Letter}$ Corresponding authors.}

\begin{abstract}
Generating coherent and diverse human dances from music signals has gained tremendous progress in animating virtual avatars. 
While existing methods support direct dance synthesis, they fail to recognize that enabling users to edit dance movements is far more practical in real-world choreography scenarios.
Moreover, the lack of high-quality dance datasets incorporating iterative editing also limits addressing this challenge.
To achieve this goal, we first construct \textbf{DanceRemix}, a large-scale multi-turn editable dance dataset comprising the prompt featuring over 25.3M dance frames and 84.5K pairs.
In addition, we propose a novel framework for iterative and editable dance generation coherently aligned with given music signals, namely \textbf{DanceEditor}. 
Considering the dance motion should be both musical rhythmic and enable iterative editing by user descriptions, our framework is built upon a prediction-then-editing paradigm unifying multi-modal conditions.
At the initial prediction stage, our framework improves the authority of generated results by directly modeling dance movements from tailored, aligned music.
Moreover, at the subsequent iterative editing stages, we incorporate text descriptions as conditioning information to draw the editable results through a  specifically designed \textbf{Cross-modality Editing Module (CEM)}.
Specifically, CEM adaptively integrates the initial prediction with music and text prompts as temporal motion cues to guide the synthesized sequences.
Thereby, the results display music harmonics while preserving fine-grained semantic alignment with text descriptions.
Extensive experiments demonstrate that our method outperforms the state-of-the-art models on our newly collected DanceRemix dataset.
Code is available at \url{https://lzvsdy.github.io/DanceEditor/}.

\end{abstract}

\vspace{-5mm}
\section{Introduction}
\label{sec:intro}
The music-conditioned dance generation seeks to create coherent and diverse human dance movements synchronizing with musical rhythms. 
These non-verbal body languages transcend cultural boundaries to communicate emotions, ideas, and narratives in real life~\cite{desmond1997meaning,  lamothe2019dancing}.
Meanwhile, modeling music-driven dance movements has broad real-world applications, including choreography creation~\cite{chen2021choreomaster, yun2014development,
qi2024cocogesture}, embodied AI~\cite{peng2015robotic, qi2023diverse}, avatar animation~\cite{stergiou2022exploring, akbas2022virtual,qi2025comathbfgesture}, human-machine interaction~\cite{stergiou2022exploring, qi2024emotiongesture,
qi2024weakly}, and virtual/augmented reality (AR/VR)~\cite{chan2019everybody, iqbal2022acceptance}.
% , and has extended to neuroscience research~\cite{kshtriya2015dance}.
In recent years, many studies have been conducted to tackle this complex challenge.

While significant progress has been made in directly music-driven dance generation~\cite{li2021ai, chen2021choreomaster, tseng2023edge, li2023finedance, luo2024popdg}, dance editing remains rarely explored. 
% They mostly overlook affording editable dance movements for users, which is more practical  in real-world scenes.
% Besides, rather than editing dance motions, others~\cite{gong2023tm2d, luo2024m, yang2024unimumo, ling2024motionllama}  utilize text and music jointly to guide dance generation by aligning music, text, and dance representations in a shared latent space.
Although some works synthesize the controllable dance sequences by incorporating text prompts and music jointly~\cite{gong2023tm2d, luo2024m, yang2024unimumo, ling2024motionllama}, they mostly overlook producing iterative high-quality results that follow the user guidance in practice.
% However, few researchers focus on developing high-quality dance datasets, including iterative editing.
Meanwhile, few researchers have devoted themselves to constructing datasets with multi-turn editable dance movements aligned with music conditions. 
For example, the choreographer may intend to iteratively edit the unsatisfactory generated initial dance by concise edit descriptions such as `lift right knee higher'' or `kick left leg twice''.
% For example, we may want to iteratively edit the unsatisfactory generated dance by concise edit descriptions, such as ``put right hand on the floor'', ``kick left leg twice'', ``lift right knee higher''.
In this work, we therefore introduce the new task of iterative editable dance generation conditioned on music and text descriptions, as shown in Figure~\ref{fig:fig_teaser}.

There are two main challenges in this task: 
% 1) High-quality dance datasets incorporating iterative editing are scarce.
% Creating such a dataset containing edit descriptions as well as source and target dance motions is difficult due to the subjective nature of dance, making it difficult to describe, combined with the costly labor required for annotations.
1) High-quality dance datasets supporting multi-turn editing are scarce. 
Creating such a dataset composed of accurate text descriptions aligned with both iterative motions and music is quite difficult. The intrinsic subjective nature of dance makes it difficult to describe, combined with the costly labor required for annotations.
2) Modeling coherent and diverse dance motions aligned with musical rhythms and tempos is challenging, especially when utilizing iteratively open-vocabulary edit descriptions.

\begin{figure}[t]
  \centering
  \includegraphics[width=0.9 \linewidth]
  {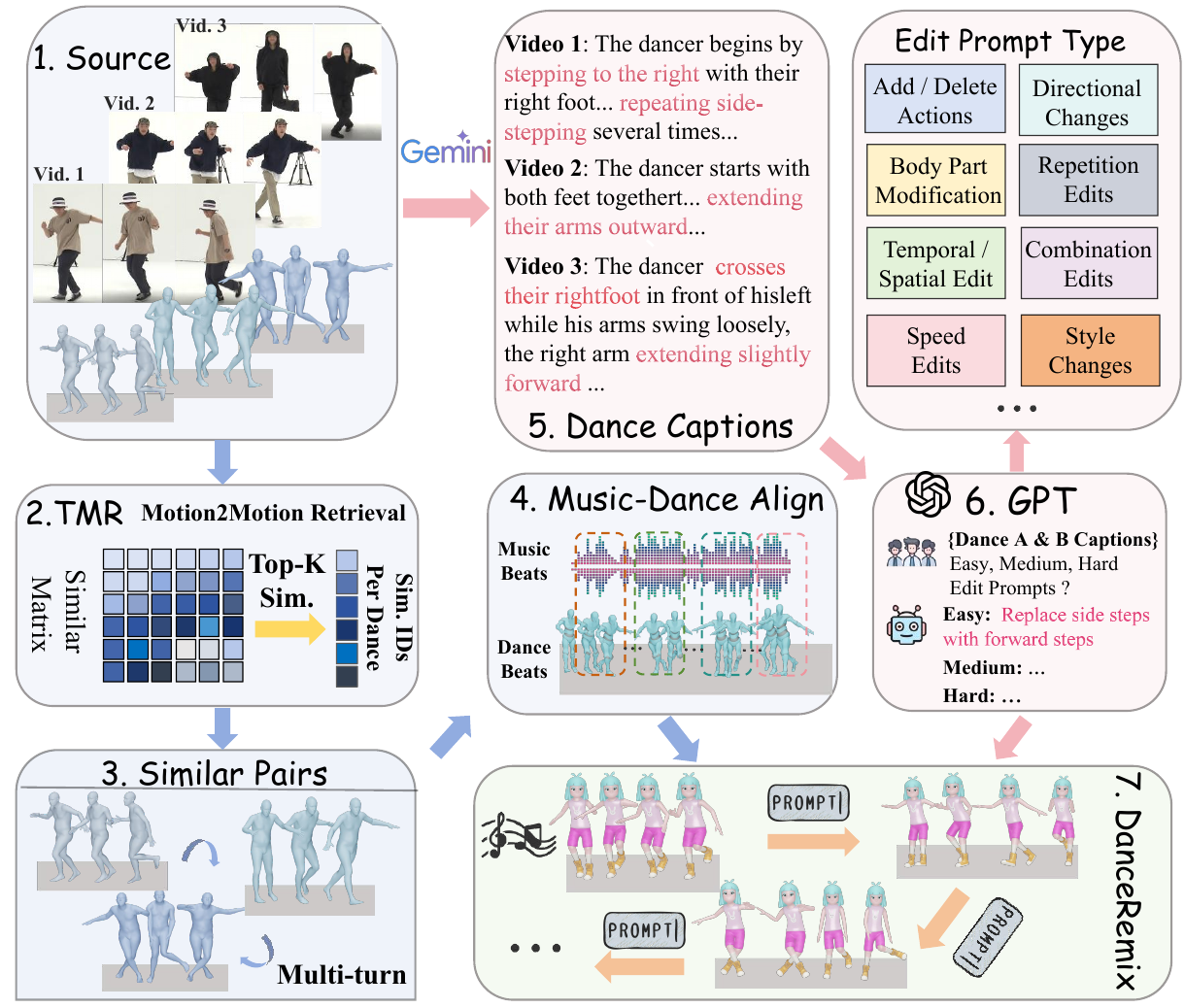}
  \caption{The workflow of DanceRemix dataset construction. 
  %get sim. motion pairs
  Firstly, we perform motion-to-motion retrieval to obtain similar dance motion pairs. 
  % align beats
  Then, we align the motion beats of the edited dance with the music beats.
  %dance caption
  For aligned dance pairs, we use Gemini to generate dense dance captions for the dance videos.
  %edit prompts
  Next, based on the generated captions, we leverage ChatGPT to generate edit instructions. 
  %multi-turn 
  Through several motion pair retrievals, we obtain music, seed dance, a series of edit prompts, and corresponding edited dance motions. 
  %datasets
  % In this way, we construct the first large-scale, multi-modal, multi-turn editable dance dataset.
  }
  % \caption{The workflow of DanceRemix dataset construction. Our initial step involves leveraging Gemini-1.5-Pro to generate dense dance captions for dance videos from existing datasets. Next, we employ TMR~\cite{petrovich2023tmr} to retrieve similar dance pairs, which are then aligned with tailored music. With the assistance of ChatGPT-4, we generate multi-grained edit prompts based on the dance captions for these similar dance pairs. By integrating music, initial dance, edit prompts, and the corresponding edited dance, we construct the first large-scale, multi-modal, multi-turn editable dance dataset.}
  \label{fig:fig_dataset_workflow}
  \vspace{-1mm}
\end{figure}

To address the issue of data scarcity, we construct a new large-scale multi-turn editable dance dataset comprising the prompt featuring over 25.3M dance frames and 84.5K pairs, dubbed \textbf{DanceRemix}, as shown in Figure~\ref{fig:fig_dataset_workflow}.
The key challenge is how to carefully build the iterative dance motions aligned with the given music signals.
Inspired by the text-motion retrieval approach TMR \cite{petrovich2023tmr}, our insight is to leverage the pre-annotated text descriptions of dance motions for constructing pairs of similar dances. 
In particular, we first employ the TMR to conduct motion-to-motion retrieval, allowing us to obtain pairs of candidates for multiple rounds of editing.
To ensure the retrieved dance motion rhythmically aligned with music beats, we use dynamic time warping to match motion and music beats as inspired by \cite{yang2024unimumo}.
Once the motion pairs are acquired, we utilize the advanced MLLM \cite{team2023gemini} to derive dense descriptions of dance movements.
Furthermore, based on these captions, we leverage ChatGPT~\cite{ouyang2022training} to produce the coherent transformation scripts from given sequences to similar ones.

Based on our DanceRemix dataset, we propose a novel framework, named \textbf{DanceEditor}, to model music-driven dance generation and enable iterative editing by open-vocabulary descriptions. 
% The key sight of our framework is to carefully build the interaction between the text and source dance motions. 
% Here, we propose a \textbf{Predicition-then-Editing Paradigm} unifying multi-modal conditions. 
% improve the authority of generated results by directly modeling dance movements from tailored aligned music.
% In the second stage, we incorporate textual descriptions, music, and initial dance predictions as conditional guidance to draw the editable results, using a ControlNet~\cite{zhang2023adding} architecture.
To achieve this goal, our framework is built to unify multi-modal conditions by a prediction-then-editing paradigm.
In the initial prediction stage, we introduce a transformer-based diffusion branch to directly produce the high-fidelity motion sequences according to tailored aligned music. 
The second stage aims to synthesize the multi-turn iterative edited motions,  followed by the guidance of the text prompt and initial results.

% Moreover, to ensure the edited dance movements are fine-grained semantic alignment with edit prompts, we devise a specifically designed Cross-modality Editing Module (\textbf{CEM}). 
To ensure the edited dance movements are fine-grained semantic alignment with edit prompts while preserving the rhythmic of the initial prediction with music signals, we present an editing paradigm called Cross-modality Editing Module (CEM), analogous to the famous ControlNet~\cite{zhang2023adding}.
% Specifically, CEM adaptively integrates the initial dance predictions with music and text prompts as temporal motion cues to guide the synthesized sequences. 
% First, we utilize a self-learning module to expand the text embedding across the temporal dimension. 
Specifically, CEM adaptively integrates the initial dance predictions with music and text prompts into a shared joint latent space.
Then, we model the joint embedding of text and initial dance, as well as text and the current iterative motions, while thoroughly considering the temporal correlation between the text and dance. 
% Here, the learned joint embedding is leveraged as a soft weight to balance the interaction dependence of the generated current dance movements on the initial prediction. 
% Here, the learned joint embedding is leveraged as a soft weight to balance the dependence of the generated current edited dance movements \wrt the initial ones. 
% Finally, we employ self-attention combined with an adaptive instance normalization (AdaIN) layer~\cite{yang2022adaint} to incorporate music into the previously fused dance movements, thereby further enhancing high-fidelity dance generation that aligns well with the edited descriptions.
The learned joint embedding is leveraged as a soft weight to provide the temporal dependence of the generated current edited dance movements \wrt the initial ones. 
Here, we infuse it into the editing branch via an adaptive instance normalization (AdaIN) layer~\cite{yang2022adaint}, thereby further enhancing high-fidelity dance generation that aligns well with the edited descriptions.
Extensive experiments on our newly collected DanceEditor dataset verify the effectiveness of our method, demonstrating diverse and plausible dance generation and editing.

Overall, our contributions are summarized as follows:
\begin{itemize}[leftmargin=*]
    \item We introduce the new task of iterative, editable music-driven dance generation with one newly collected large-scale dataset named DanceRemix, significantly paving the way for diverse dance generation and editing.

    \item We propose DanceEditor, a novel framework that enables high-fidelity multi-turn iterative choreography following a prediction-then-editing paradigm.

    \item We present a custom-designed Cross-modality Editing Module (CEM) to encourage the temporal synchronization of dance movements \wrt multi-modal conditions, including music, and text prompts, thereby allowing open-vocabulary text-guided coherent dance motion generation.

    \item Extensive experiments show that our method outperforms state-of-the-art counterparts, displaying realistic and impressive iterative editable dance sequence generation.

\end{itemize}

\section{Related Work}

% one linewidth

\textbf{Music to Dance Generation:} 
Early studies in music-to-dance synthesis focused on template-based approaches, including example-based alignment ~\cite{fan2011example}, statistical music-motion mapping ~\cite{ofli2011learn2dance}, and music similarity-driven motion stitching ~\cite{lee2013music}. While pioneering, these methods suffer from short motion duration, limited diversity, and unnatural transitions due to reliance on handcrafted rules.
Subsequent deep learning approaches improve generalization: GrooveNet ~\cite{alemi2017groovenet} and ChoreoMaster ~\cite{chen2021choreomaster} leverage neural networks for real-time synthesis, while LSTM-autoencoders ~\cite{tang2018dance} and adversarial learning frameworks like DeepDance ~\cite{sun2020deepdance} enhance temporal coherence. 
Nevertheless, deterministic generation paradigms lead to rigid motions and restricted style diversity.

The advent of Transformers revolutionizes temporal modeling. Works like AI choreographer ~\cite{li2021ai} and Music2Dance ~\cite{zhuang2022music2dance} utilize cross-modal attention to align music and motion features, achieving superior rhythmic synchronization. However, challenges like unnatural joint rotations and motion discontinuities persist in a complex choreography. Recent generative models further advance the field: GAN-based methods ~\cite{kim2022brand} improve continuity but face mode collapse, while diffusion models (\eg, EDGE ~\cite{tseng2023edge}, Bailando ~\cite{zhang2024bidirectional}) achieve state-of-the-art quality through iterative denoising. Despite progress, critical limitations remain high computational costs, temporal inconsistency in long sequences, and an inability to support iterative user editing, which is essential for practical choreography.

\noindent \textbf{Text to Motion Generation:}
Text-to-motion research primarily follows two paradigms. Align-based methods ~\cite{petrovich2022temos, guo2022generating, tevet2022motionclip} project text and motion into shared latent spaces but struggle with fine-grained correlations. Condition-based models like MotionDiffuse ~\cite{zhang2022motiondiffuse}, MDM ~\cite{Tevet2022HumanMD}, and MLD ~\cite{chen2023executing} directly inject text features into generators via diffusion, yet neglect editability. Recent editing techniques attempt to bridge this gap: FineMoGen ~\cite{zhang2024finemogen} edits spatiotemporal details but requires restrictive input formats. MotionFix ~\cite{athanasiou2024motionfix} leverages paired data but fails in dance-specific edits. MotionLab ~\cite{guo2025motionlab} unifies generation/editing but lacks whole-body control. A key limitation across these works is the disruption of music-dance synchronization during editing which is also limited by data scarcity.
Existing datasets (\eg, AIST++ ~\cite{aist-dance-db}, FineDance ~\cite{li2023finedance}) focus on single-turn generation, lacking edit trajectories or textual instructions. This impedes the development of frameworks requiring iterative refinement. Our work addresses these gaps through the DanceRemix dataset and a multi-conditioned (music + text) editing framework, enabling coherent and user-controllable dance synthesis.

\label{sec:related}

\section{Proposed Method}

\begin{figure*}[t]
  \centering
  \includegraphics[width=0.9\textwidth]{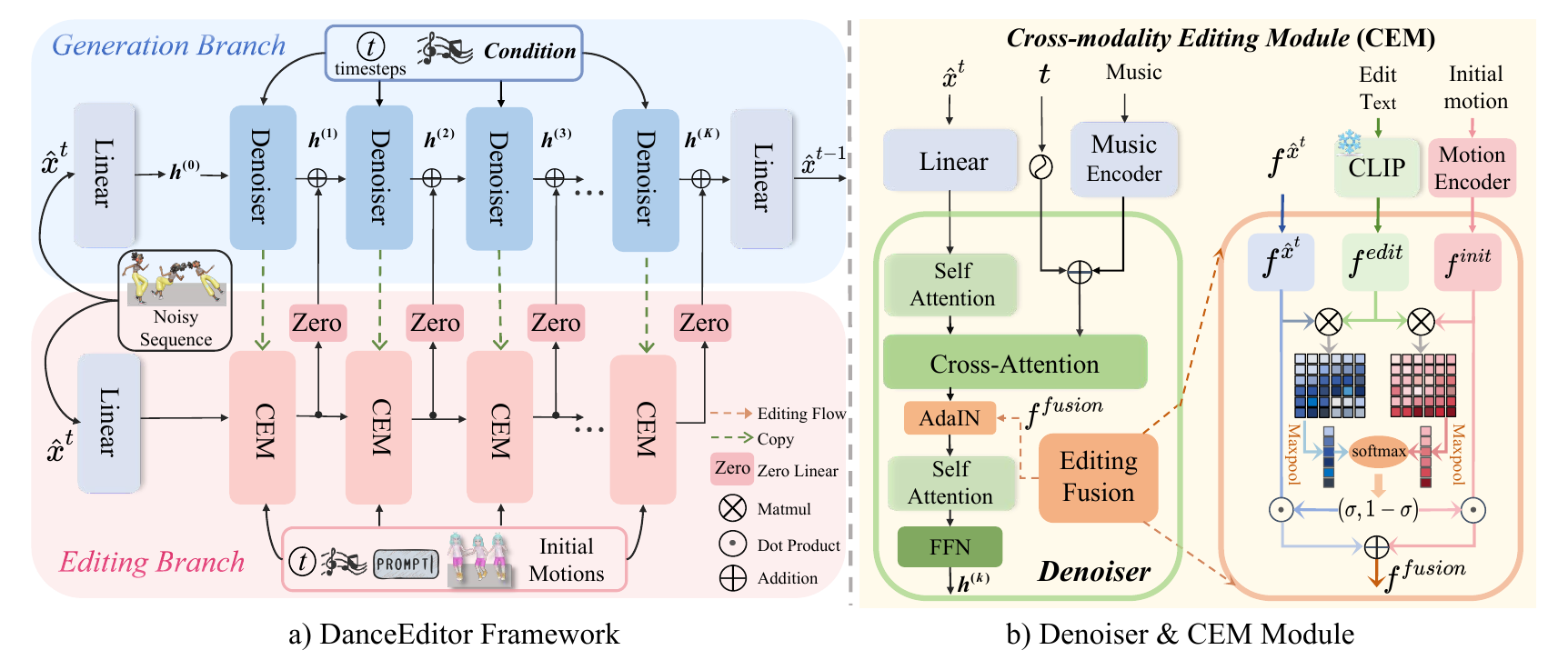}
  \caption{
  In the initial prediction stage, a diffusion transformer-based  \textbf{Generating Branch} takes music signals as input and synthesizes vivid dance motions. During the second stage, the \textbf{Editing Branch} that contains a \textbf{Cross-modality Editing Module (CEM)} adaptively incorporates the initial dance predictions with both the music and text prompts, guiding the generation of edited dance sequences.
  % (b) The CEM module incorporates the hidden feature $f^{\hat{x}^t}$, extracted from the cross-attention layer of the denoiser block, along with the edit text
  % and the initial dance predictions to yield embeddings ${f^{\hat{x}^t}}$, $f^{edit}$ and $f^{init}$.
  % These embeddings are utilized to compute the correlation matrix that guides the dance fusion. Finally, the results of the dance fusion are incorporated into the denoiser block through $f^{fusion}$ via AdaIN.
  \vspace{-2mm}
  }
%   \caption{(a) \textbf{Generating Branch}: This branch produces vivid, diverse dances that correspond to the provided musical rhythm, based on conditional music input. \textbf{Editing Branch}: This branch iteratively refines initial dance results by incorporating music, the initial dance, and textual edit prompts. 
%   The edited dances align with the textual instructions while preserving rhythmic similarity to the original dance.
%   (b) \textbf{CEM module}, based on the Denoiser module, adaptively integrates initial dance predictions with music and textual prompts within a unified latent space. 
%   It leverages temporal correlations between the edit prompt and the initial dance, as well as between the edit prompt and current iterative motions, ensuring refined dance generation while maintaining coherence with the initial dance.
% }
  \label{fig: fig_framework}
\end{figure*}

\subsection{Iterative Editable Dance Dataset Construction}

% 我们有 text 有 music 有 iterative 有 motion transformation scripts
% 主要体现我们的edit prompt，IMP，其他数据集没有
% Intro: our insight is to leverage the pre-annotated text descriptions of dance motions for constructing pairs of similar dances
%数据集属性 特色
%to the best of our knowledge 大， large ， firstm 支持哪些任务
% 对齐的motions和edit dance
% 我们的亮点是什么，music(s)的意思放到最后一句解释
% lz TODO 要解释 edit prompt 单位，要不然confuse，可能让别人以为是vocabulary

\begin{table}[t]
\centering
\setlength{\tabcolsep}{0.2cm} % 调整列间距
\caption{
% Statistical comparisons of our \textbf{DanceRemix} dataset against various counterparts. 
To the best of our knowledge, our \textbf{DanceRemix} is the first large-scale dataset that enables \textbf{iterative} editable dance generation. Given each music segment, our dataset contains at least two different iterable editing sequence pairs (\ie, $\times$ 2) with corresponding editing prompts. 
% DanceRemix is the first large-scale multimodal dance dataset featuring well-aligned music segments ("music(s)"), seamlessly connected multi-segment dance motions, and fine-grained edit prompts.
% Our dataset enables iterative editable generation while addressing data scarcity in dance research.
}
\footnotesize
\begin{tabular}{lcccc}
    \toprule
    \textbf{Dataset} & \textbf{Edit prompt} & \textbf{Motion} & \textbf{Music(hours)} \\ 
    \midrule \midrule
    % Dance with Melody\cite{ref1} & $\times$ &  & 5640 \\ 
    Duolando\cite{siyao2024duolando} & \ding{55} & 334 & 0.38 \\ 
    PopDanceSet\cite{luo2024popdg} & \ding{55} & 1,036 & 0.56 \\ 
    Finedance \cite{li2023finedance} & \ding{55} & 211 & 14.6 \\  
    AIST++\cite{li2021ai} & \ding{55} & 1,408 & 5.19 \\ 
    \midrule
    \rowcolor[HTML]{ECF4FF} \textbf{DanceRemix(Ours)} & \textbf{84,523} & \textbf{84,523$\times$2} & \textbf{117.39} \\ 
    \bottomrule
\end{tabular}
\label{tab:DanceRemix_stas}
\end{table}

% 为什么构造数据集的亮点 ！！
% caption和edit都很好
%难点：没有对齐的音乐，edit，多段舞蹈，舞蹈之间的衔接编辑， transformation scripts
\noindent\textbf{Dataset Description:}
% motivation => 缓解数据集的缺少
%出于什么目的，构建了什么数据集，scale， 多模态
%数据集亮点： 大(frames, pairs,时长)， 多个模态完美align到一块, 为什么要有真多属性，原因是什么
%为了缓解数据缺乏性==> 构建了xx数据集，包含xxx frames, pairs,  hours
%due to the limited modality attributes in the existing dance generation dataset, we newly collect  DanceRemix, upon prior works~\cite{li2021ai}  %放大别人缺点，突出我们的有点
% To be  specific, 我们数据集 规模 
% 我们数据量大==> 模型充分建模， 包含更加多样的人体动作， 有利于去生成 vivid and xxx dance motions
%Additionnally, to support our insight on iterative editable dance generation, our dataset has well-aligned motions and edit prompts.
%具体而言，这些代表什么。(别人没有，我们有的)
%我们文本well aligned, 同一个音乐对应多段舞蹈和音乐合拍，多段舞蹈之间细粒度的transformation scripts (i.e. edit prompts)
% Due to the limited single  in the existing dance generation dataset, we newly collect DanceRemix, upon prior works~\cite{li2021ai}.
Due to the deficiency of single mapping of dance and music in existing datasets, we newly collect DanceRemix for iterative editable dance generation, upon prior works~\cite{li2021ai, luo2024popdg, lin2023motion}.
To be precise, 
This extensive dataset ensures diverse dance movements, allowing our model to learn intricate motion patterns and generate vivid dance motions.
Additionally, to support our insight into iterative editable dance generation, our dataset has well-aligned dance motions and edit prompts.
Specifically, several dance sequences corresponding to the same music piece are well-synchronized with musical rhythms and beats. 
The transitions among these dance segments are seamlessly aligned with the edit prompts.
To the best of our knowledge, this is the first large-scale multi-turn editable dance dataset, as reported in Table~\ref{tab:DanceRemix_stas}.

\noindent\textbf{Automatic Data Collection Workflow:}
Constructing datasets conducive to our task focuses on formulating one-to-many mapping (one music signal \wrt several iterative movement pairs), and synthesising fine-grained editing text prompts among pairs of movements. 
% Leveraging advanced MLLMs to generate captions and transformation scripts for similar dance pairs is both practical and valuable, given the inherently subjective nature of dance, which makes it difficult to describe.
Thanks to advanced large multi-modality model approaches, we devise an automatic data collection workflow as illustrated in Figure~\ref{fig:fig_dataset_workflow}.
To obtain pairs of similar dance sequences, we employ a text-motion retrieval approach, TMR ~\cite{petrovich2023tmr},  to perform motion-to-motion retrieval based on re-organized motion sequences from prior datasets. 
Then we apply a top-k selection strategy to identify the similar pairs with an obvious but natural transition. 
To ensure the retrieved dance motions are rhythmically aligned with the same piece of music, we use dynamic time warping~\cite{yang2024unimumo} to determine the optimal alignment between musical and motion beats. 
In this manner, movements that are not in harmony with the beat of the music are dropped out and manually checked. 

Once several similar dance sequences well-aligned with the same music piece are acquired, we use Gemini \cite{team2023gemini} to generate fine-grained motion captions. 
Based on these captions, we leverage ChatGPT \cite{ouyang2022training} to create coherent transformation scripts representing editing text prompts from one sequence to another similar one.
Notably, the richness and detail of the captions enable diverse edit prompts, including additions, deletions, body part modifications, and temporal or spatial edits, as shown in Figure~\ref{fig:fig_dataset_workflow}.
By integrating these methods, we can automatically and cost-effectively generate a large-scale, multi-turn editable dance dataset with well-aligned motions and edit prompts.

\subsection{Problem Formulation}
%预先定义很多后面会用到的名称以及相关字符
%规定好输入和输出
%规定好rep方式
%不要在描述方法来添加过多的字符，用具体物理意义来阐释
%eg. good:  jukebox feature， we extract audio feature as xxx guidance to lead the generation of x xxxx
% bad: we use the encoder  to extract the feature fa to add xxx
% 在正文的方法中，尽量减少字符的定义
% 图里面有字符 ，正文一定有
% 对于具体比如矩阵相乘，softmax
% framework, two-staged input and output

% Our proposed DanceEditor framework is designed to synthesize realistic dance sequences, which can subsequently be stylized through a later iterative editing process. 
Given a sequence of music signal $M=\{m_1,\dots,m_N\}$, our DancerEditor framework aims to produce iterative editable human dance movements $X=\{x_1,\dots,x_N\}$ with the aid of open-vocabulary descriptions.
Here, $N$ represents the total frame length of dance sequences.
For dance posture $x_i$ in each frame, we leverage the 24-joint SMPL~\cite{loper2023smpl} format to represent body structure, where each joint is represented in 6D rotation representation for better modeling.
Besides, we introduce a 3-dimensional root position and a 4-dimensional binary foot contact to indicate the global body movements in a continuous sequence.
Similar to the prior work~\cite{tseng2023edge}, we utilize the advanced audio technique Jukebox~\cite{dhariwal2020jukebox} to map the raw music waves into rhythmic features.

\subsection{Prediction-then-Editing Paradigm}

Considering the iterative editable dance movement synthesis should preserve coherent alignment with both the given music signals and editing text prompts, we present a novel prediction-then-editing paradigm to unify the incorporation of these multimodal conditions.
As shown in Figure~\ref{fig: fig_framework}, the prediction-then-editing paradigm is collaboratively composed of an initial prediction stage and an iterative editing stage.
In the initial prediction stage, we utilize a music-conditioned dance diffusion model to synthesize dance movements from tailored, aligned music.
During the iterative editing stage, we intend to exploit the text prompts as guidance for editable dance motion production.

\subsubsection{Music-conditioned Initial Prediction}
\label{sec: 3.3}
%geneation branch 重要性
% The transformer-based diffusion branch within our DanceEditor framework is the foundation for high-quality, iteratively editable dance generation. 
% In the initial prediction stage, as illustrated in Figure~\ref{fig: fig_framework}, the Generating Branch, Music-Conditioned Dance Diffusion, is a transformer-based diffusion model that also serves as the foundation for subsequent high-quality, iteratively editable dance generation.
As illustrated in Figure~\ref{fig: fig_framework}, the initial prediction stage is composed of a diffusion transformer-based generation branch.
This branch is capable of effectively learning both the beat and tempo of the dance, thus ensuring synchronization and alignment with the accompanying music signals.
%main branch 目标
% This branch is capable of effectively learning both the beat and tempo of the dance, thus ensuring synchronization and alignment with the accompanying music signals.
%main branch 输入输出
During training, our denoiser is trained to generate continuous dance motions from noisy dance sequences $x^t$, conditioned on the diffusion timesteps $t$ and the music condition $c_m$.
We first feed the noisy dance sequences into a linear layer to obtain the hidden state of encoded dance movement, indicated as $h^{k}$, where $k \in \left \{ 0,...,K \right \} $ is the stacked denoiser blocks for $K$ times.
The denoising process is guided by the simple objective:
\begin{align}
\mathcal{L}_{simple} & = \mathbb{E} _{\mathrm{x},t,c_m,\epsilon \sim \mathcal{N}(0, 1) }  \left [\left \| \mathrm{x}-\mathcal{D}_{c}(\mathrm{x}^{t},t,c_m)\right \|_{2}^{2}    \right ],
\label{eq1}
\end{align}
where $\mathcal{D}_{c}$ represents our music-conditional denoiser, $\epsilon\sim \mathcal{N}(\mathbf{0},\mathbf{I})$ is the added random Gaussian noise, and $\mathrm{x}^{t} = \mathrm{x} + \sigma _{t}\epsilon$ is the gradually noise adding process at step $t$. 
Here, $\sigma _{t} \in (0, 1)$ is a constant hyperparameter.
Moreover, following the settings of ~\cite{tevet2022human, guo2022generating}, we adopt the velocity loss $\mathcal{L}_{vel}$ and foot contact loss $\mathcal{L}_{foot}$ to improve the smoothness and physical plausibility of the generated dance motions. Accordingly, the overall objective is defined as:
\begin{align}
\mathcal{L}_{total} & = \lambda _{simple} \mathcal{L}_{simple} + \mathcal{L}_{vel} +  \mathcal{L}_{foot},
\label{eq2}
\end{align}
where $\lambda _{simple}$ is trade-off weight coefficients.

\subsubsection{Iterative Editable Dance Generation}
%editable dance genartion输入
% During the second editing stage, we refine the Music-Conditioned Dance Diffusion model by adding control conditions, namely edit descriptions and the initial dance results.
% To preserve the robust music-driven dance generation capabilities of the generating branch, we introduce the Cross-modality Editing Module (CEM), an editing paradigm analogous to ControlNet. 
% This module allows the editing branch to be fine-tuned, leveraging the knowledge acquired by the frozen generating branch, without directly modifying the generating branch.
In the following editing stage, we incorporate the editing transformation prompts\footnotetext[1]{Here, the editing transformation prompts usually represent the movement changes from the initial sequence to the target one, such as ``lifting your right leg higher'', or ``swing your arms wider''.} to achieve iterative editable dancer generation. 
Inspired by ControlNet~\cite{zhang2023adding}, we introduce an editing branch to inject the condition signals into the dance motion generation, as depicted in Figure~\ref{fig: fig_framework}.

\noindent\textbf{Cross-modality Editing Module:}
\label{module: CEM}
% To effectively capture the relevance of temporal edits from edit prompts $c_{edit}$ to iteratively edited dance motions $\hat{x}_{edit}$, while preserving rhythmic alignment between the initial dance motions $x_{init}$ and music signals$c_{m}$, we propose CEM, which adaptively facilitates these interactions.
% We extract the hidden feature embedding $f_h^{\hat{x}_t}$ of $\hat{x}_{edit}$ which has incorporates both the temporal information of the dance sequences and the interaction with music signals, achieved through an attention mechanismanism.
% To enable effective feature fusion, we project the features representing edit text $c_{edit}$ , initial dance $c_{init}$ and current iterative motions $f_h^{\hat{x}_t}$ into the same latent space, obtaining embeddings  $f^{edit}$, $f^{init}$ , and $f^{\hat{x}_t}$, respectively.
To ensure the edited dance movements have fine-grained semantic alignment with edit prompts while preserving the rhythm of the initial prediction with music signals, we design a Cross-modality Editing Module (CEM). 
Considering that the editing prompts usually display as the temporal-wise effects upon the final results, our CEM effectively captures the temporal relevance between the extracted text prompt features and the motion embeddings of iteratively edited dance.
In particular, we first include the addition of a denoising timestep and music condition as the query feature to match the key feature and value feature belonging to the current noisy motion via a cross-attention mechanism, thereby obtaining the music-coherent current sequence embedding.

% Then, we compute the temporal correlation matrix $M^{init}$ between the embeddings of edit text $f^{edit}$ and init dance $f^{init}$, and the matrix $M^{edit}$ between edit text $f^{edit}$ and current iterative motion $f_h^{\hat{x}_t}$. 
Then, to achieve the fine-grained semantic control of the updated current sequence embedding, we introduce an editing fusion block. 
Here, we calculate the temporal correlation matrix between the updated current dancer embedding and extracted text embedding as $M^{edit}\in \mathbb{R}^{N\times N}$. 
This matrix represents the temporal variants of the editing impact upon the current dance sequence.
Subsequently, we conduct a similar operation to acquire the matrix that denotes the temporal deformation of the editing impact upon the initial dance sequence, as $M^{init}\in \mathbb{R}^{N\times N} $.
% The correlation matrix $M^{init}\in \mathbb{R}^{N\times N} $ and $M^{edit}$ is passed through maxpooling to obtain the dominant correlating embeddings $e^{init}$ and $e^{edit}$, as
% \begin{equation}
% \begin{aligned}
% e^{init} &= \operatorname{MaxPool} \left( f^{init} \otimes f^{edit} \right), \\
% e^{edit} &= \operatorname{MaxPool} \left( f^{init} \otimes f_h^{\hat{x}_t} \right).
% \end{aligned}
% \label{eq3}
% \end{equation}
Furthermore, we obtain the two learnable parameters represented by dance fusion weights of the initial dances and the current iterative dance motions, as follows
\begin{align}
(\sigma, 1-\sigma)=\operatorname{Softmax} (AdPool(M^{edit}), AdPool(M^{init})),
\label{eq3}
\end{align}
where $AdPool$ indicates the AdaptiveMaxPooling operation along the temporal dimension.
% $M^{edit} = f^{\hat{x}_t} \otimes f^{edit}$

% Once we acquire these weight parameters, the current fusion dance embedding is boosted as follows:
Once we acquire these weight parameters, we leverage them to obtain the fused motion embedding by adaptively integrating the current motion sequence features and the initial ones, formulated as:
\begin{align}
f^{fusion} =  \sigma  \cdot  f^{init}  + (1-\sigma) \cdot f^{\hat{x}_t},
\label{eq4}
\end{align}
where $f^{init}$ denotes the initial motion features and $f^{\hat{x}_t}$ indicates the current iterative movement features. 
Through this fashion, the obtained motion fusion embedding contains the collaborative guidance representation with the help of edit cues. 
The collaborative guidance representation is then exploited to boost the current iterative motion features by an adaptive instance normalization (AdaIN) layer~\cite{yang2022adaint}.
% The embedding $f^{fusion}$, which represents the adaptive fusion of the current denoised motions and the initial dance motions under the guidance of the edit cues, is then added to the original latent motion embedding $f_h^{\hat{x}_t}$ through AdaIN~\cite{yang2022adaint}, refining it to $f'^{\hat{x}_t}_{h}$, which is represented by

\begin{align}
% {f^{\prime \hat{x}_t}_h}= \operatorname{AdaIN}(f^{\hat{x}_t}, f^{fusion}),
{f'^{\hat{x}_t}_h}= \operatorname{AdaIN}(f^{\hat{x}_t}, f^{fusion}),
\label{eq5}
\end{align}
where ${f^{\prime \hat{x}_t}_h}$ represents edited motion embedding. In this fashion, CEM allows the editing branch to effectively integrate temporal motion cues from the initial dance movements and edit prompts, enabling the iterative generation of editable dances through prompt adjustments. For the iterative editing stage of text prompt conditioned dance generation, our training objective is consistent with the previous initial prediction stage.
\section{Experiments}

\begin{table*}[t]
\centering
\caption{Comparison of our DanceEditor framework and the state-of-the-art methods on our DanceRemix dataset and POPDG dataset. 
    $\uparrow$ denotes the higher the better, and $\downarrow$ indicates the lower the better.}
  \vspace{-0.5em}
\footnotesize
\setlength{\tabcolsep}{3.5 mm}
\begin{tabular}{lcccccccc}
\toprule
\multirow{2}{*}{Models} & \multicolumn{4}{c}{DanceRemix Dataset} & \multicolumn{4}{c}{POPDG Dataset}  \\ \cmidrule(r){2-5}  \cmidrule(r){6-9}
                        & FID $\downarrow$ & BAS $\uparrow$ & Diversity $\uparrow$ & PFC $\downarrow$ & FID $\downarrow$ & BAS $\uparrow$ & Diversity $\uparrow$ & PFC $\downarrow$  \\ \midrule \midrule
EDGE~\cite{tseng2023edge}\textcolor[HTML]{C0C0C0}{$_{CVPR'23}$}                & 3.91   & 0.2519  & 2.29  & 1.635 &   4.26 &  0.2491  & 2.14 & 6.156 \\
TM2D~\cite{gong2023tm2d}\textcolor[HTML]{C0C0C0}{$_{ICCV'23}$}                 & 3.84  & 0.2470 &  2.16 & 1.327 &  4.15  & 0.2466  & 2.09 & 5.340\\ 
Lodge ~\cite{li2024lodge}\textcolor[HTML]{C0C0C0}{$_{CVPR'24}$}                & 3.57  & 0.2545 & 2.92  &  1.559 & 4.03  & 0.2473 & 2.57 &  5.788\\
POPDG~\cite{luo2024popdg}\textcolor[HTML]{C0C0C0}{$_{CVPR'24}$}              & 4.02  & 0.2513  & 2.64  & 1.122 & 4.32  & 0.2502 & 2.33 & 4.875\\ 
 \midrule

\rowcolor[HTML]{ECF4FF} 
\textbf{DanceEditor} & \textbf{2.83} & \textbf{0.2560} & \textbf{3.12} & \textbf{0.784} & \textbf{2.87} 
        & \textbf{0.2535} &\textbf{3.05} &\textbf{3.514}\\ 
\bottomrule
\end{tabular}
\label{tab:tab_compare}
\vspace{-1em}
\end{table*}

\subsection{Datasets and Experimental Setting}
\label{sec: 4.1}
%数据集怎么收集
%数据集包含了啥
\noindent\textbf{DanceRemix Dataset:} 
%当前数据集不行，我们构建了 DanceRemix to evaluate our approach
Since the existing dance generation datasets fail to provide accurate edit descriptions aligned with both dance motions and music, we contribute a new dataset named DanceRemix to evaluate our approach, upon prior work~\cite{li2021ai, luo2024popdg, lin2023motion}.
%数据集怎么构建
The dance motions of our DanceRemix are collected from online sources and existing dance datasets. We then employ TMR to retrieve similar dance pairs, ensuring their alignment with music beats through dynamic time warping. To generate coherent and natural transformation scripts for motions in these similar dance pairs, we leverage the state-of-the-art MLLM\cite{team2023gemini} to extract dense captions from dance videos. Subsequently, we utilize ChatGPT\cite{ouyang2022training} to generate diverse editing instructions. 
This data collection workflow takes more than three months, acquiring over 84.5 K motion pairs with corresponding music/edit descriptions.
To further advance multi-turn editable dance generation, we extend DanceRemix to DanceRemix-X. By transforming dance pairs, we also construct three-level edit descriptions representing different granularities.
For more details, please refer to the supplementary materials.

% 数据集
% 1. 当前数据集的问题是什么，对应 abstract 和 intro
% 2. 我们的数据集从哪来的(简单介绍)
% 3. 数据集规模\内容\参数(待统计,不完全)

% \noindent\textbf{DanceRemix Dataset:} To address the challenge of data scarcity in multi-turn dance editing, we construct \textit{DanceRemix}. This large-scale dataset includes 12.6 million dance frames organized into 42K high-quality prompt-motion pairs. Each pair encapsulates multi-turn editable dance sequences generated through a rigorous pipeline of motion capture and semantic alignment. The dataset spans diverse dance genres and temporal granularities, with sequence lengths 5 seconds at 60 FPS.

% 训练细节
%  1. 总述我们用的模型
%  2. 两个阶段的训练参数,因为是diffusion,所以我认为应该谈到推理速度的问题.但是咱们的这个任务又不care推理速度
%  3. wav 特征的对比,第五章可以提到我们的对比实验
% ps: 参考 freemotion和edge

%1. 输入输出, 特征细节
%2. 训练阶段细节(我们独有的细节）
% 
\noindent\textbf{Implementation Details:} 
%s数据setting
We set the motion and music sequence length to 5 seconds, namely $N=150$ frames. Additionally, we apply first-frame canonicalization to ensure that the motions face the same direction and have identical initial global positions, similar to the approaches in~\cite{petrovich2022temos, athanasiou2024motionfix}.
Temporally, our DanceEditor synthesizes the 5-second dance motions including $24$ joints in practice. Each joint is converted to the 6D rotation representation~\cite{zhou2019continuity} for better modeling in the experiments.

%实验setting
In the initial prediction stage, we set $\lambda _{simple} = 10$, empirically. The total diffusion time step is $1,000$ with the cosine noisy schedule~\cite{nichol2021improved}. The initial learning rate is set as $1\times 10^{-4}$ with AdamW optimizer.
Our model is trained on $8$ NVIDIA H800 GPUs with a batch size of $128$.  The total training process takes a total of $250$ epochs, accounting for 1 day. 
During the following editing stage, we train the CEM module with a batch size of 96 for 200 epochs. The initial learning rate is set as $1\times 10^{-5}$. We take the DDIM~\cite{song2020denoising} sampling strategy with $50$ denoising timesteps for inference.

\noindent\textbf{Evaluation Metrics:}

\begin{itemize}
    \item \textbf{FID:} We compute the Fréchet Inception Distance (FID) using features extracted from an independently trained motion autoencoder, following previous works~\cite{chen2021choreomaster, li2023finedance, ling2024motionllama}. A lower FID indicates that the generated dance motion better matches the ground truth distribution.
    
    \item \textbf{BAS:} The Beat Alignment Score (BAS) measures the temporal alignment between music and dance by calculating the correspondence of music beats with sequence movements. The higher score means the better music-dance rhythmic alignment.

    \item \textbf{Diversity:} The Diversity score measures the distance between generated motion features conditioned on different music inputs. The higher score means the better diversity of generated results.
    \item \textbf{PFC:} The Physical Foot Contact (PFC) score evaluates the plausibility of dance movements based on hip acceleration and foot velocity, where a lower score signifies more realistic foot-ground contact.
    
    \item \textbf{Motion-Editing Text Align Score (MEAS):} Inspired by ~\cite{liang2024intergen}, we measure the distance between dance motion pairs and editing texts using a custom-trained CLIP-based model.  A lower score indicates better alignment between the edited dance and the editing descriptions.

    % user study  text-edited-dance align? user study our methods win

\end{itemize}

\subsection{Quantitative Evaluation}
\noindent{\textbf{Comparisons with SOTA Methods:}}
To the best of our knowledge, we are the first to explore iterative,  editable music-driven dance generation with open-vocabulary descriptions. 
%music-to-dance
To thoroughly validate the superiority of our method, we retrain various state-of-the-art (SOTA) Music-to-Dance \textbf{(M2D)} approaches on our DanceRemix dataset setting. For diffusion-based methods—EDGE~\cite{tseng2023edge}, Lodge~\cite{li2024lodge}, and POPDG~\cite{luo2024popdg}—as well as VQ-VAE models like TM2D~\cite{gong2023tm2d}, we leverage the official implementations and adhere to the published hyperparameter configurations.
% % advantage of iterative editable dance generation over direct generation based on text and music
% Additionally, for music-plus-text-to-motion \textbf{(MT2D)} task, we explore adapting these M2D methods to incorporate edit texts alongside music to jointly guide dance generation, highlighting the advantages of iterative editable dance generation over directly generating dances conditioned on edit texts and music. 
% Specifically, for each comparison method, we modify text embeddings from the final layer of CLIP to match the dimensions and add them to music and diffusion timesteps to facilitate motion generation. More results are included in Appendix

As shown in Table~\ref{tab:tab_compare}, we adopt the FID, BAS, Diversity, and PFC for a well-rounded view of comparison. Our DanceEditor outperforms all the competitors by a large margin on the DanceRemix dataset. Remarkably, our method even achieves more than 21\% ($i.e., (3.57  -  2.83 ) / 3.57 \approx 21\%$) improvement over the sub-optimal counterpart in FID. 
% Diversity
We observe that both Lodge~\cite{li2024lodge} and ours synthesize the authority dance motions with much higher diversity  than others. 
This is because both benefit from the two-stage paradigms, which separately focus on music-aligned global dance patterns and highly effective local dance quality.
However, Lodge shows lower performance on FID due to its lack of learning fine-grained dance transformations from given sequences to similar ones.
%BC eval
In terms of BAS, our method achieves much better results than other counterparts. 
This strongly aligns with our insight into the prediction-then-editing paradigm and CEM, which encourages our model to further learn the local variations of temporal motions and their interactions with music beats and rhythms. 

Additionally, our method achieves a significantly lower PFC compared to others, indicating more physically realistic motions.

We further evaluate our method on the POPDG dataset, with the results presented in Table~\ref{tab:tab_compare}. These findings further validate the superiority of our proposed method.
% % 额外实验介绍
% Additionally, for the multi-conditioned (Music + Text) dance generation task, we conducted further comparative experiments by adapting M2D methods to integrate edit texts alongside music. For more details, please refer to supplementary materials.

\begin{table}[t]
  \caption{Ablation on Prediction-then-Editing Paradigm. $\uparrow$ denotes the higher the better, and $\downarrow$ indicates the lower the better. }
  \vspace{-0.5em}
  \centering
    \footnotesize
    \setlength{\tabcolsep}{1.5 mm}
    \begin{tabular}{lccccc}
    \toprule
    \multirow{2}{*}{Models} & \multicolumn{4}{c}{DanceRemix Dataset} \\ \cmidrule(r){2-5} 
                            & FID $\downarrow$ & BAS $\uparrow$ & Diversity $\uparrow$  & MEAS $\downarrow$ \\ \midrule \midrule
    Ours (Initial)   & \textbf{2.83} & \textbf{0.2560} &  3.12  &  \ding{55}\\
    Ours (Iteration \#1)   & 2.85  & 0.2553  &  3.16& \textbf{0.784}\\ 
    Ours (Iteration \#2)  & 2.91 & 0.2541 & 3.23 &  0.786\\ 
    Ours (Iteration \#3)   & 3.04 & 0.2524 & \textbf{3.35} &  0.793\\

    \bottomrule
    \end{tabular}
  \label{tab:tab_paradigm_ablation}
\end{table}

% \begin{table}[t]
%   \vspace{-1em}
%   \caption{Ablation of editing branch (\textbf{1-round editing}) \&\& Comparison between direct (combining three prompts into one) and multi-turn editing. $\uparrow$ denotes the higher the better, and $\downarrow$ indicates the lower the better. }
%   \vspace{-0.5em}
%   \centering
%     \footnotesize
%     \setlength{\tabcolsep}{1.4 mm}
%     \begin{tabular}{lccccc}
%     \toprule
%     \multirow{2}{*}{Models} & \multicolumn{4}{c}{DanceRemix Dataset} \\ \cmidrule(r){2-5} 
%                             & FID $\downarrow$ & BAS $\uparrow$ & Diversity $\uparrow$  & MEAS $\downarrow$ \\ \midrule \midrule
%      w/o Editing Brach   & 3.95  & 0.2514 & 2.32 &   1.351\\
%      Editing Branch w/o CEM   & 3.68   & 0.2537 & 2.69  & 1.024 \\
%      Editing Branch w/o inital motion  & 3.51   &  0.2548 & 2.74  & 0.988\\
%      Editing Branch w/o music           & 3.34   & 0.2539  & 2.71  &  0.932\\
     
%     \rowcolor[HTML]{ECF4FF} 
%     \textbf{DanceEditor (1-round editing)}  & \textbf{2.85}  & \textbf{0.2553} & \textbf{3.16}  & \textbf{0.784} \\ 
%     % \midrule
%     \hline
%     direct editing                      &  3.21          & 0.2519         &  2.92           &  0.925     \\
%     \rowcolor[HTML]{FFF9DB} 
%     \textbf{3-round editing}            & \textbf{3.04}  & \textbf{0.2524} & \textbf{3.35}  & \textqbf{0.793} \\
%     % \bottomrule
%     \hline
%     \end{tabular}
%     \vspace{-1.5em}
    
%   \label{tab:tab_module_ablation}
% \end{table}
\begin{table}[t]
  \caption{Ablation on Cross-modaliy Editing Module. $\uparrow$ denotes the higher the better, and $\downarrow$ indicates the lower the better. }
   \vspace{-0.5em}
  \centering
    \footnotesize
    \setlength{\tabcolsep}{1.5 mm}
    \begin{tabular}{lccccc}
    \toprule
    \multirow{2}{*}{Models} & \multicolumn{4}{c}{DanceRemix Dataset} \\ \cmidrule(r){2-5} 
                            & FID $\downarrow$ & BAS $\uparrow$ & Diversity $\uparrow$  & MEAS $\downarrow$ \\ \midrule \midrule
     w/o Editing Brach   & 3.95  & 0.2514 & 2.32 &   1.351\\
     Editing Branch w/o CEM   & 3.68   & 0.2537 & 2.69  & 1.024 \\
     \midrule
    \rowcolor[HTML]{ECF4FF} 
    \textbf{DanceEditor (full version)}  & \textbf{2.85}  & \textbf{0.2553} & \textbf{3.16}  & \textbf{0.784} \\ 
    \bottomrule
    \end{tabular}
  \label{tab:tab_module_ablation}
   \vspace{-0.5em}
\end{table}

\begin{figure*}[t]
	\centering  %origin 16.6 cm
	\includegraphics[width=0.98\textwidth]{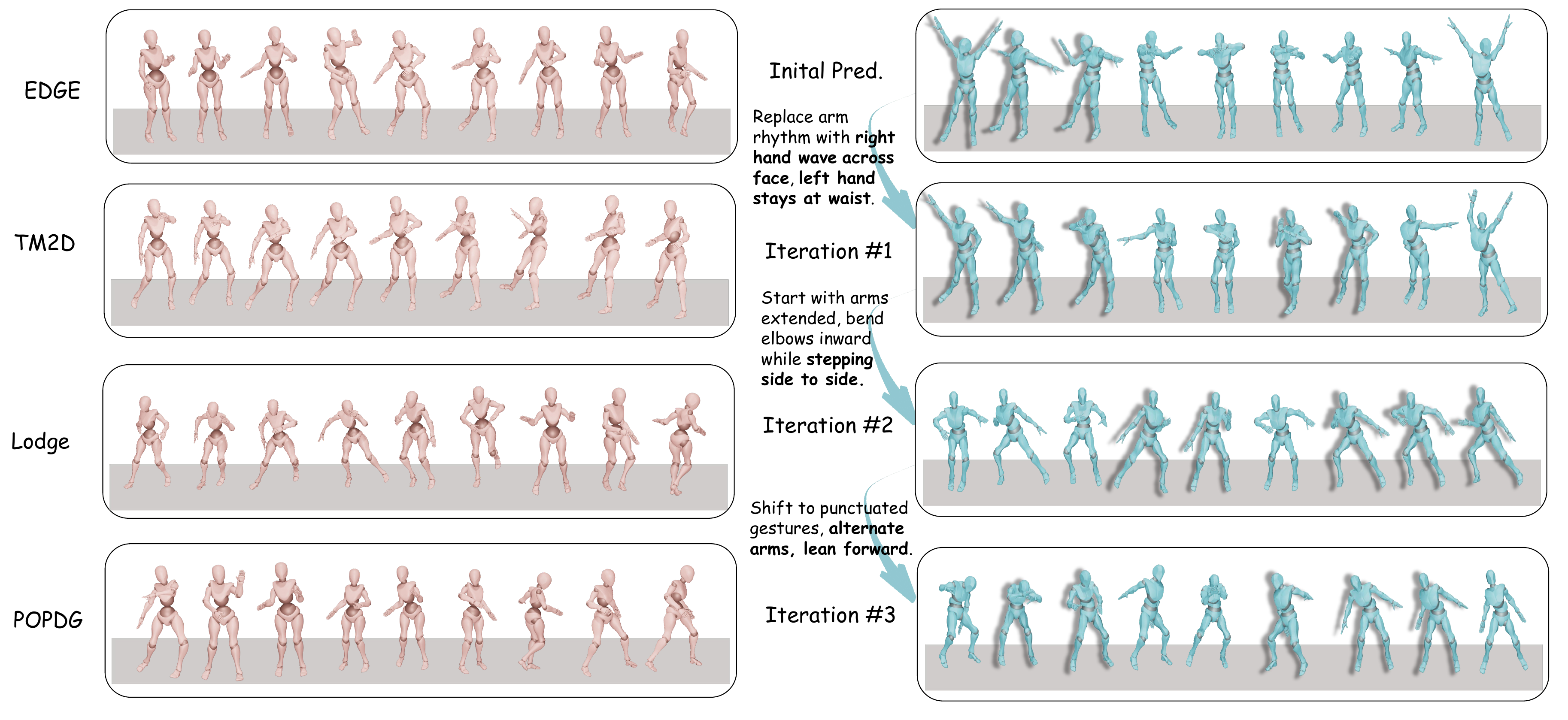}
    \vspace{-0.5em} 
	\caption{Given the same music segments, the results generated by our \textbf{DanceEditor} framework and other SOTA comparison methods.}
    \vspace{-1.3em} 
	\label{fig:fig_vis}
\end{figure*}

\noindent\textbf{Ablation Study:} Here, to evaluate the effectiveness of our proposed prediction-then-editing paradigm and CEM Module, we conduct extensive ablation studies separately.

\noindent\textbf{Ablation on Prediction-then-Editing Paradigm:} 
To verify the effectiveness of our proposed prediction-then-editing paradigm, we formulate the ablation on multi-turn iterative editing, and the results are reported in Table~\ref{tab:tab_paradigm_ablation}. Given a period of music signals, our framework enables the generation of high-quality sequences while preserving harmonic rhythm with the beat of the music. The initial prediction is produced directly by the generation branch from music signals. Then, we integrate the initial prediction with the corresponding music \&  text editing prompts into the editing branch to synthesize the editable dance movements (\ie, denoted as Iteration \#1). The subsequent iterations are generated in a similar way.
The results demonstrate that our prediction-then-editing paradigm effectively promotes the framework to model the dance movements in a carefully learned joint distribution space. 

In particular, with continuous iterative editing, the performance of our model always tends to a stable state. The metrics Diversity clearly attain much better results than the initial ones. This indicates that the editing text prompts in our dataset significantly facilitate the framework to learn prolific dynamic features. 
The MEAS metric slightly decreases due to minor motion shifts from the previous editing iteration, but the overall visual quality remains excellent.
Although the FID displays slightly worse scores than the initial ones due to the sampled open-vocabulary text descriptions, our method still achieves the optimal result. Meanwhile, we observe that the BAS scores still perform better than the state-of-the-art counterparts. This applies the effectiveness of our editing paradigm that can adaptively model the temporal dependency between the music beats and iterative dance movements. Our prediction-then-editing paradigm demonstrates a superior ability to generate iterative editable dance motions with a unified framework.

\noindent\textbf{Ablation on the CEM module:} To further validate the effectiveness of our Cross-Modality Editing Module (CEM), we conduct ablation studies to assess the necessity of the editing branch and the effectiveness of CEM. 
Specifically, we evaluate experimental settings where both music and editing text are used as inputs to generate dance motions.
As reported in ~\ref{tab:tab_module_ablation}, we design three models for comparison: (1)  only the generation branch, (2) two branches without the CEM module, and (3) DanceEditor with both prediction and editing. The first two models use simple concatenation to fuse music and textual features.
To be specific, the generation-only branch struggles to directly learn the complex interactions between dance motions, music, and text. As a result, implementing the model without an editing branch leads to worse performance across all metrics.
Meanwhile, our CEM brings significant improvements across all metrics compared to the editing branch without CEM. This demonstrates that CEM effectively enhances the temporal dependency between music beats and iterative dance movements, leading to vivid and coherent motions.

\subsection{Qualitative Evaluation}

\begin{figure}[t]
  \centering
    % \begin{center}
    \includegraphics[width=0.95\linewidth]{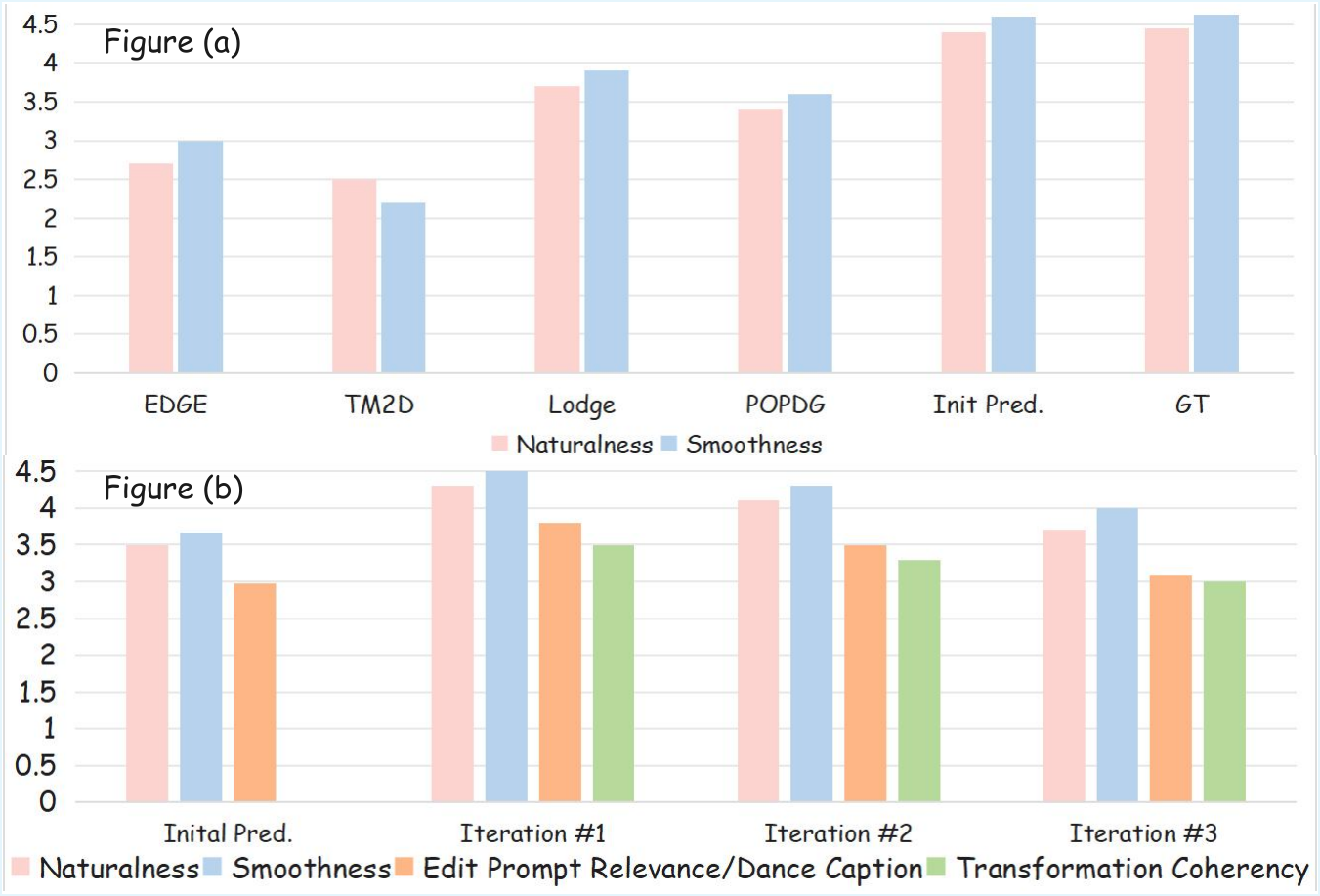}
    % \end{center}
    \vspace{-0.5em}
    
  % \begin{subfigure}[b]{1\linewidth}
  %   \includegraphics[width=1\linewidth]{figure/usr_study1.png}
  %   \caption{}
  %   \label{fig:user_study_a}
  % \end{subfigure}
  
  % \begin{subfigure}[b]{1\linewidth}
  %   \includegraphics[width=1\linewidth]{figure/usr_study2.png}
  %   \caption{}
  %   \label{fig:user_study_b}
  % \end{subfigure}
  
\caption{User study with dance naturalness, motion smoothness, Motion-Editing Text Alignment, and motion transformation coherency on our dataset. }
\vspace{-2em} 
   \label{fig:user_study}
\end{figure}

\noindent\textbf{Visualization:}
To showcase the superior performance of our method, we present visualized keyframes from our DanceEditor framework alongside other approaches on our DanceRemix dataset.
Unlike others, which generate dance solely from music, our method showcases both the initial prediction and iteratively edited dance motions
% 分析
As shown in Figure~\ref{fig:fig_vis}, our method generates vivid and diverse dance movements compared to others. Specifically, EDGE and TM2D tend to produce stiff and unnatural results. While Lodge generates relatively varied motions, it suffers from foot drifting and abrupt changes. POPDG, though more natural, lacks expressiveness.
In contrast, our DanceRemix not only generates natural and expressive dance movements but also seamlessly edits motions while maintaining coherence with the edit instructions, aligning well with our prediction-then-editing paradigm.
For more visualization demos, please refer to the videos available on our project website.
% In our experiments, the generated dance sequences are 150 frames long at 30 FPS, ensuring all user study demo videos have a uniform duration of 5 seconds.

\noindent\textbf{User Study:}
To assess the quality of synthesized dance movements, we conducted a user study with 15 anonymously recruited participants from diverse academic backgrounds.
For each comparison method and our prediction stage, we randomly selected two generated videos. Additionally, for our iterative editing results, we presented two editing processes, each with an initial prediction and three iterations.
Each participant should watch 18 videos, each lasting 5 seconds. They rated the dance results for the M2D task based on naturalness and motion smoothness. For our iterative editable dance generation, we introduced two additional evaluation criteria: edit prompt relevance and transformation coherence. 
The visualized videos are randomly selected, ensuring at least two samples per method. The results, shown in Figure \ref{fig:user_study}, are rated on a scale of 0 to 5 (higher is better).
Our framework outperforms all competitors, particularly excelling in smoothness, highlighting the effectiveness of our prediction-then-editing paradigm. 
Even after multiple editing iterations, edit prompt relevance and transformation coherence decline only slightly, with CEM ensuring motions remain well-aligned with the edit texts.

\section{Conclusion}
%我们提出xxxx框架，xxx的工作，实现什么效果，具体包含module的功能是什么，实现了什么目的
%limitation和 
%future work sentence 
In this paper, we introduce DanceEditor, a novel framework for iterative and editable dance generation coherently aligned with given music signals. 
To fulfill this goal, we first newly collect a large-scale multi-turn editable dance dataset, DanceRemix,  featuring over 25.3M dance frames and 84.5K pairs.
Along with this dataset, our approach generates vivid dance motions and enables coherent editing guided by open-vocabulary descriptions.
To further capture the temporal synchronization of dance movements \wrt multimodal conditions, we propose a Cross-modality Editing Module that adaptively integrates the initial dance predictions with both music and texts to guide edited dance genreation. 
Extensive experiments conducted on DanceRemix show the superiority of our model.

\noindent \textbf{Limitation} 
Currently, we support only the generation and editing of body movements in dance sequences.
In the future, we will explore semantic descriptions and editing of facial expressions and gestures to create more expressive and realistic dance performances coherent with music.

% \vspace{0.5em}
\noindent{\textbf{Acknowledgements.}} This work is funded in part by the HKUST Start-up Fund (R9911), Theme-based Research Scheme grant (T45-205/21-N), the InnoHK funding for Hong Kong Generative AI Research and Development Center, Hong Kong SAR, and the research funding under HKUST-DXM AI for Finance Joint Laboratory (DXM25EG01).

% \newpage

% \newpage
% \appendix
% \input{sec/X_suppl}

{\small
    \bibliographystyle{ieeenat_fullname}
    \bibliography{main}

\begin{thebibliography}{56}
\providecommand{\natexlab}[1]{#1}
\providecommand{\url}[1]{\texttt{#1}}
\expandafter\ifx\csname urlstyle\endcsname\relax
  \providecommand{\doi}[1]{doi: #1}\else
  \providecommand{\doi}{doi: \begingroup \urlstyle{rm}\Url}\fi

\bibitem[Akbas et~al.(2022)Akbas, Evren~Yantac, Eskenazi, Kuscu, Semsioglu, Topal~Sumer, and Ozturk]{akbas2022virtual}
Saliha Akbas, Asim Evren~Yantac, Terry Eskenazi, Kemal Kuscu, Sinem Semsioglu, Onur Topal~Sumer, and Asli Ozturk.
\newblock Virtual dance mirror: A functional approach to avatar representation through movement in immersive vr.
\newblock In \emph{Proceedings of the 8th International Conference on Movement and Computing}, pages 1--4, 2022.

\bibitem[Alemi et~al.(2017)Alemi, Fran{\c{c}}oise, and Pasquier]{alemi2017groovenet}
Omid Alemi, Jules Fran{\c{c}}oise, and Philippe Pasquier.
\newblock Groovenet: Real-time music-driven dance movement generation using artificial neural networks.
\newblock \emph{networks}, 8\penalty0 (17):\penalty0 26, 2017.

\bibitem[Athanasiou et~al.(2024)Athanasiou, Cseke, Diomataris, Black, and Varol]{athanasiou2024motionfix}
Nikos Athanasiou, Alp{\'a}r Cseke, Markos Diomataris, Michael~J Black, and G{\"u}l Varol.
\newblock Motionfix: Text-driven 3d human motion editing.
\newblock In \emph{SIGGRAPH Asia 2024 Conference Papers}, pages 1--11, 2024.

\bibitem[Chan et~al.(2019)Chan, Ginosar, Zhou, and Efros]{chan2019everybody}
Caroline Chan, Shiry Ginosar, Tinghui Zhou, and Alexei~A Efros.
\newblock Everybody dance now.
\newblock In \emph{Proceedings of the IEEE/CVF international conference on computer vision}, pages 5933--5942, 2019.

\bibitem[Chen et~al.(2021)Chen, Tan, Lei, Zhang, Guo, Zhang, and Hu]{chen2021choreomaster}
Kang Chen, Zhipeng Tan, Jin Lei, Song-Hai Zhang, Yuan-Chen Guo, Weidong Zhang, and Shi-Min Hu.
\newblock Choreomaster: choreography-oriented music-driven dance synthesis.
\newblock \emph{ACM Transactions on Graphics (TOG)}, 40\penalty0 (4):\penalty0 1--13, 2021.

\bibitem[Chen et~al.(2023)Chen, Jiang, Liu, Huang, Fu, Chen, and Yu]{chen2023executing}
Xin Chen, Biao Jiang, Wen Liu, Zilong Huang, Bin Fu, Tao Chen, and Gang Yu.
\newblock Executing your commands via motion diffusion in latent space.
\newblock In \emph{Proceedings of the IEEE/CVF Conference on Computer Vision and Pattern Recognition}, pages 18000--18010, 2023.

\bibitem[Desmond(1997)]{desmond1997meaning}
Jane Desmond.
\newblock \emph{Meaning in motion: New cultural studies of dance}.
\newblock Duke University Press, 1997.

\bibitem[Dhariwal et~al.(2020)Dhariwal, Jun, Payne, Kim, Radford, and Sutskever]{dhariwal2020jukebox}
Prafulla Dhariwal, Heewoo Jun, Christine Payne, Jong~Wook Kim, Alec Radford, and Ilya Sutskever.
\newblock Jukebox: A generative model for music.
\newblock \emph{arXiv preprint arXiv:2005.00341}, 2020.

\bibitem[Fan et~al.(2011)Fan, Xu, and Geng]{fan2011example}
Rukun Fan, Songhua Xu, and Weidong Geng.
\newblock Example-based automatic music-driven conventional dance motion synthesis.
\newblock \emph{IEEE transactions on visualization and computer graphics}, 18\penalty0 (3):\penalty0 501--515, 2011.

\bibitem[Gong et~al.(2023)Gong, Lian, Chang, Guo, Jiang, Zuo, Mi, and Wang]{gong2023tm2d}
Kehong Gong, Dongze Lian, Heng Chang, Chuan Guo, Zihang Jiang, Xinxin Zuo, Michael~Bi Mi, and Xinchao Wang.
\newblock Tm2d: Bimodality driven 3d dance generation via music-text integration.
\newblock In \emph{Proceedings of the IEEE/CVF International Conference on Computer Vision}, pages 9942--9952, 2023.

\bibitem[Guo et~al.(2022)Guo, Zou, Zuo, Wang, Ji, Li, and Cheng]{guo2022generating}
Chuan Guo, Shihao Zou, Xinxin Zuo, Sen Wang, Wei Ji, Xingyu Li, and Li Cheng.
\newblock Generating diverse and natural 3d human motions from text.
\newblock In \emph{Proceedings of the IEEE/CVF conference on computer vision and pattern recognition}, pages 5152--5161, 2022.

\bibitem[Guo et~al.(2025)Guo, Hu, Zhao, and Soh]{guo2025motionlab}
Ziyan Guo, Zeyu Hu, Na Zhao, and De~Wen Soh.
\newblock Motionlab: Unified human motion generation and editing via the motion-condition-motion paradigm.
\newblock \emph{arXiv preprint arXiv:2502.02358}, 2025.

\bibitem[Iqbal and Sidhu(2022)]{iqbal2022acceptance}
Javid Iqbal and Manjit~Singh Sidhu.
\newblock Acceptance of dance training system based on augmented reality and technology acceptance model (tam).
\newblock \emph{Virtual Reality}, 26\penalty0 (1):\penalty0 33--54, 2022.

\bibitem[Kim et~al.(2022)Kim, Oh, Kim, Tong, and Lee]{kim2022brand}
Jinwoo Kim, Heeseok Oh, Seongjean Kim, Hoseok Tong, and Sanghoon Lee.
\newblock A brand new dance partner: Music-conditioned pluralistic dancing controlled by multiple dance genres.
\newblock In \emph{Proceedings of the IEEE/CVF Conference on Computer Vision and Pattern Recognition}, pages 3490--3500, 2022.

\bibitem[LaMothe(2019)]{lamothe2019dancing}
Kimerer LaMothe.
\newblock The dancing species: how moving together in time helps make us human.
\newblock \emph{Aeon, June}, 1\penalty0 (1):\penalty0 1--2, 2019.

\bibitem[Lee et~al.(2013)Lee, Lee, and Park]{lee2013music}
Minho Lee, Kyogu Lee, and Jaeheung Park.
\newblock Music similarity-based approach to generating dance motion sequence.
\newblock \emph{Multimedia tools and applications}, 62:\penalty0 895--912, 2013.

\bibitem[Li et~al.(2021)Li, Yang, Ross, and Kanazawa]{li2021ai}
Ruilong Li, Shan Yang, David~A Ross, and Angjoo Kanazawa.
\newblock Ai choreographer: Music conditioned 3d dance generation with aist++.
\newblock In \emph{Proceedings of the IEEE/CVF International Conference on Computer Vision}, pages 13401--13412, 2021.

\bibitem[Li et~al.(2023)Li, Zhao, Zhang, Su, Ren, Zhang, Tang, and Li]{li2023finedance}
Ronghui Li, Junfan Zhao, Yachao Zhang, Mingyang Su, Zeping Ren, Han Zhang, Yansong Tang, and Xiu Li.
\newblock Finedance: A fine-grained choreography dataset for 3d full body dance generation.
\newblock In \emph{Proceedings of the IEEE/CVF International Conference on Computer Vision}, pages 10234--10243, 2023.

\bibitem[Li et~al.(2024)Li, Zhang, Zhang, Zhang, Guo, Zhang, Liu, and Li]{li2024lodge}
Ronghui Li, YuXiang Zhang, Yachao Zhang, Hongwen Zhang, Jie Guo, Yan Zhang, Yebin Liu, and Xiu Li.
\newblock Lodge: A coarse to fine diffusion network for long dance generation guided by the characteristic dance primitives.
\newblock In \emph{Proceedings of the IEEE/CVF Conference on Computer Vision and Pattern Recognition}, pages 1524--1534, 2024.

\bibitem[Liang et~al.(2024)Liang, Zhang, Li, Yu, and Xu]{liang2024intergen}
Han Liang, Wenqian Zhang, Wenxuan Li, Jingyi Yu, and Lan Xu.
\newblock Intergen: Diffusion-based multi-human motion generation under complex interactions.
\newblock \emph{International Journal of Computer Vision}, 132\penalty0 (9):\penalty0 3463--3483, 2024.

\bibitem[Lin et~al.(2023)Lin, Zeng, Lu, Cai, Zhang, Wang, and Zhang]{lin2023motion}
Jing Lin, Ailing Zeng, Shunlin Lu, Yuanhao Cai, Ruimao Zhang, Haoqian Wang, and Lei Zhang.
\newblock Motion-x: A large-scale 3d expressive whole-body human motion dataset.
\newblock \emph{Advances in Neural Information Processing Systems}, 36:\penalty0 25268--25280, 2023.

\bibitem[Ling et~al.(2024)Ling, Han, Li, Shen, Cheng, and Zou]{ling2024motionllama}
Zeyu Ling, Bo Han, Shiyang Li, Hongdeng Shen, Jikang Cheng, and Changqing Zou.
\newblock Motionllama: A unified framework for motion synthesis and comprehension.
\newblock \emph{arXiv preprint arXiv:2411.17335}, 2024.

\bibitem[Loper et~al.(2023)Loper, Mahmood, Romero, Pons-Moll, and Black]{loper2023smpl}
Matthew Loper, Naureen Mahmood, Javier Romero, Gerard Pons-Moll, and Michael~J Black.
\newblock Smpl: A skinned multi-person linear model.
\newblock In \emph{Seminal Graphics Papers: Pushing the Boundaries, Volume 2}, pages 851--866. 2023.

\bibitem[Luo et~al.(2024{\natexlab{a}})Luo, Hou, Li, Chang, Liu, Wang, and Shan]{luo2024m}
Mingshuang Luo, Ruibing Hou, Zhuo Li, Hong Chang, Zimo Liu, Yaowei Wang, and Shiguang Shan.
\newblock M$^{3}$ gpt: An advanced multimodal, multitask framework for motion comprehension and generation.
\newblock \emph{arXiv preprint arXiv:2405.16273}, 2024{\natexlab{a}}.

\bibitem[Luo et~al.(2024{\natexlab{b}})Luo, Ren, Hu, Huang, and Yao]{luo2024popdg}
Zhenye Luo, Min Ren, Xuecai Hu, Yongzhen Huang, and Li Yao.
\newblock Popdg: Popular 3d dance generation with popdanceset.
\newblock In \emph{Proceedings of the IEEE/CVF Conference on Computer Vision and Pattern Recognition}, pages 26984--26993, 2024{\natexlab{b}}.

\bibitem[Nichol and Dhariwal(2021)]{nichol2021improved}
Alexander~Quinn Nichol and Prafulla Dhariwal.
\newblock Improved denoising diffusion probabilistic models.
\newblock In \emph{International conference on machine learning}, pages 8162--8171. PMLR, 2021.

\bibitem[Ofli et~al.(2011)Ofli, Erzin, Yemez, and Tekalp]{ofli2011learn2dance}
Ferda Ofli, Engin Erzin, Y{\"u}cel Yemez, and A~Murat Tekalp.
\newblock Learn2dance: Learning statistical music-to-dance mappings for choreography synthesis.
\newblock \emph{IEEE Transactions on Multimedia}, 14\penalty0 (3):\penalty0 747--759, 2011.

\bibitem[Ouyang et~al.(2022)Ouyang, Wu, Jiang, Almeida, Wainwright, Mishkin, Zhang, Agarwal, Slama, Ray, et~al.]{ouyang2022training}
Long Ouyang, Jeffrey Wu, Xu Jiang, Diogo Almeida, Carroll Wainwright, Pamela Mishkin, Chong Zhang, Sandhini Agarwal, Katarina Slama, Alex Ray, et~al.
\newblock Training language models to follow instructions with human feedback.
\newblock \emph{Advances in neural information processing systems}, 35:\penalty0 27730--27744, 2022.

\bibitem[Peng et~al.(2015)Peng, Zhou, Hu, Chao, and Li]{peng2015robotic}
Hua Peng, Changle Zhou, Huosheng Hu, Fei Chao, and Jing Li.
\newblock Robotic dance in social robotics—a taxonomy.
\newblock \emph{IEEE Transactions on Human-Machine Systems}, 45\penalty0 (3):\penalty0 281--293, 2015.

\bibitem[Petrovich et~al.(2022)Petrovich, Black, and Varol]{petrovich2022temos}
Mathis Petrovich, Michael~J Black, and G{\"u}l Varol.
\newblock Temos: Generating diverse human motions from textual descriptions.
\newblock In \emph{European Conference on Computer Vision}, pages 480--497. Springer, 2022.

\bibitem[Petrovich et~al.(2023)Petrovich, Black, and Varol]{petrovich2023tmr}
Mathis Petrovich, Michael~J Black, and G{\"u}l Varol.
\newblock Tmr: Text-to-motion retrieval using contrastive 3d human motion synthesis.
\newblock In \emph{Proceedings of the IEEE/CVF International Conference on Computer Vision}, pages 9488--9497, 2023.

\bibitem[Qi et~al.(2023)Qi, Liu, Sun, Li, Fan, and Yu]{qi2023diverse}
Xingqun Qi, Chen Liu, Muyi Sun, Lincheng Li, Changjie Fan, and Xin Yu.
\newblock Diverse 3d hand gesture prediction from body dynamics by bilateral hand disentanglement.
\newblock In \emph{Proceedings of the IEEE/CVF conference on computer vision and pattern recognition}, pages 4616--4626, 2023.

\bibitem[Qi et~al.(2024{\natexlab{a}})Qi, Liu, Li, Hou, Xin, and Yu]{qi2024emotiongesture}
Xingqun Qi, Chen Liu, Lincheng Li, Jie Hou, Haoran Xin, and Xin Yu.
\newblock Emotiongesture: Audio-driven diverse emotional co-speech 3d gesture generation.
\newblock \emph{IEEE Transactions on Multimedia}, 26:\penalty0 10420--10430, 2024{\natexlab{a}}.

\bibitem[Qi et~al.(2024{\natexlab{b}})Qi, Pan, Li, Yuan, Chi, Li, Luo, Xue, Zhang, Liu, et~al.]{qi2024weakly}
Xingqun Qi, Jiahao Pan, Peng Li, Ruibin Yuan, Xiaowei Chi, Mengfei Li, Wenhan Luo, Wei Xue, Shanghang Zhang, Qifeng Liu, et~al.
\newblock Weakly-supervised emotion transition learning for diverse 3d co-speech gesture generation.
\newblock In \emph{Proceedings of the IEEE/CVF Conference on Computer Vision and Pattern Recognition}, pages 10424--10434, 2024{\natexlab{b}}.

\bibitem[Qi et~al.(2024{\natexlab{c}})Qi, Zhang, Wang, Pan, Liu, Li, Chi, Li, Xue, Zhang, et~al.]{qi2024cocogesture}
Xingqun Qi, Hengyuan Zhang, Yatian Wang, Jiahao Pan, Chen Liu, Peng Li, Xiaowei Chi, Mengfei Li, Wei Xue, Shanghang Zhang, et~al.
\newblock Cocogesture: Toward coherent co-speech 3d gesture generation in the wild.
\newblock \emph{arXiv preprint arXiv:2405.16874}, 2024{\natexlab{c}}.

\bibitem[Qi et~al.(2025)Qi, Wang, Zhang, Pan, Xue, Zhang, Luo, Liu, and Guo]{qi2025comathbfgesture}
Xingqun Qi, Yatian Wang, Hengyuan Zhang, Jiahao Pan, Wei Xue, Shanghang Zhang, Wenhan Luo, Qifeng Liu, and Yike Guo.
\newblock Co$^{3}$gesture: Towards coherent concurrent co-speech 3d gesture generation with interactive diffusion.
\newblock In \emph{The Thirteenth International Conference on Learning Representations}, 2025.

\bibitem[Siyao et~al.(2024)Siyao, Gu, Yang, Lin, Liu, Ding, Yang, and Loy]{siyao2024duolando}
Li Siyao, Tianpei Gu, Zhitao Yang, Zhengyu Lin, Ziwei Liu, Henghui Ding, Lei Yang, and Chen~Change Loy.
\newblock Duolando: Follower gpt with off-policy reinforcement learning for dance accompaniment.
\newblock \emph{arXiv preprint arXiv:2403.18811}, 2024.

\bibitem[Song et~al.(2020)Song, Meng, and Ermon]{song2020denoising}
Jiaming Song, Chenlin Meng, and Stefano Ermon.
\newblock Denoising diffusion implicit models.
\newblock \emph{arXiv preprint arXiv:2010.02502}, 2020.

\bibitem[Stergiou and Vosinakis(2022)]{stergiou2022exploring}
Marina Stergiou and Spyros Vosinakis.
\newblock Exploring costume-avatar interaction in digital dance experiences.
\newblock In \emph{Proceedings of the 8th International Conference on Movement and Computing}, pages 1--6, 2022.

\bibitem[Sun et~al.(2020)Sun, Wong, Cheng, Kankanhalli, Geng, and Li]{sun2020deepdance}
Guofei Sun, Yongkang Wong, Zhiyong Cheng, Mohan~S Kankanhalli, Weidong Geng, and Xiangdong Li.
\newblock Deepdance: music-to-dance motion choreography with adversarial learning.
\newblock \emph{IEEE Transactions on Multimedia}, 23:\penalty0 497--509, 2020.

\bibitem[Tang et~al.(2018)Tang, Jia, and Mao]{tang2018dance}
Taoran Tang, Jia Jia, and Hanyang Mao.
\newblock Dance with melody: An lstm-autoencoder approach to music-oriented dance synthesis.
\newblock In \emph{Proceedings of the 26th ACM international conference on Multimedia}, pages 1598--1606, 2018.

\bibitem[Team et~al.(2023)Team, Anil, Borgeaud, Alayrac, Yu, Soricut, Schalkwyk, Dai, Hauth, Millican, et~al.]{team2023gemini}
Gemini Team, Rohan Anil, Sebastian Borgeaud, Jean-Baptiste Alayrac, Jiahui Yu, Radu Soricut, Johan Schalkwyk, Andrew~M Dai, Anja Hauth, Katie Millican, et~al.
\newblock Gemini: a family of highly capable multimodal models.
\newblock \emph{arXiv preprint arXiv:2312.11805}, 2023.

\bibitem[Tevet et~al.(2022{\natexlab{a}})Tevet, Gordon, Hertz, Bermano, and Cohen-Or]{tevet2022motionclip}
Guy Tevet, Brian Gordon, Amir Hertz, Amit~H Bermano, and Daniel Cohen-Or.
\newblock Motionclip: Exposing human motion generation to clip space.
\newblock In \emph{European Conference on Computer Vision}, pages 358--374. Springer, 2022{\natexlab{a}}.

\bibitem[Tevet et~al.(2022{\natexlab{b}})Tevet, Raab, Gordon, Shafir, Cohen-Or, and Bermano]{Tevet2022HumanMD}
Guy Tevet, Sigal Raab, Brian Gordon, Yonatan Shafir, Daniel Cohen-Or, and Amit~H. Bermano.
\newblock Human motion diffusion model.
\newblock \emph{ArXiv}, abs/2209.14916, 2022{\natexlab{b}}.

\bibitem[Tevet et~al.(2022{\natexlab{c}})Tevet, Raab, Gordon, Shafir, Cohen-or, and Bermano]{tevet2022human}
Guy Tevet, Sigal Raab, Brian Gordon, Yoni Shafir, Daniel Cohen-or, and Amit~Haim Bermano.
\newblock Human motion diffusion model.
\newblock In \emph{The Eleventh International Conference on Learning Representations}, 2022{\natexlab{c}}.

\bibitem[Tseng et~al.(2023)Tseng, Castellon, and Liu]{tseng2023edge}
Jonathan Tseng, Rodrigo Castellon, and Karen Liu.
\newblock Edge: Editable dance generation from music.
\newblock In \emph{Proceedings of the IEEE/CVF Conference on Computer Vision and Pattern Recognition}, pages 448--458, 2023.

\bibitem[Tsuchida et~al.(2019)Tsuchida, Fukayama, Hamasaki, and Goto]{aist-dance-db}
Shuhei Tsuchida, Satoru Fukayama, Masahiro Hamasaki, and Masataka Goto.
\newblock Aist dance video database: Multi-genre, multi-dancer, and multi-camera database for dance information processing.
\newblock In \emph{Proceedings of the 20th International Society for Music Information Retrieval Conference, {ISMIR} 2019}, pages 501--510, Delft, Netherlands, 2019.

\bibitem[Yang et~al.(2022)Yang, Jin, Jia, Xu, and Chen]{yang2022adaint}
Canqian Yang, Meiguang Jin, Xu Jia, Yi Xu, and Ying Chen.
\newblock Adaint: Learning adaptive intervals for 3d lookup tables on real-time image enhancement.
\newblock In \emph{Proceedings of the IEEE/CVF Conference on Computer Vision and Pattern Recognition}, pages 17522--17531, 2022.

\bibitem[Yang et~al.(2024)Yang, Su, Zhang, Chen, Qian, Liu, and Gan]{yang2024unimumo}
Han Yang, Kun Su, Yutong Zhang, Jiaben Chen, Kaizhi Qian, Gaowen Liu, and Chuang Gan.
\newblock Unimumo: Unified text, music and motion generation.
\newblock \emph{arXiv preprint arXiv:2410.04534}, 2024.

\bibitem[Yun et~al.(2014)Yun, Kim, Lee, and Lee]{yun2014development}
Hye-jeong Yun, Kwang-il Kim, Jeong-hun Lee, and Hae-Yeoun Lee.
\newblock Development of experience dance game using kinect motion capture.
\newblock \emph{KIPS transactions on software and data engineering}, 3\penalty0 (1):\penalty0 49--56, 2014.

\bibitem[Zhang et~al.(2024{\natexlab{a}})Zhang, Tang, Zhang, Lin, Han, Xiao, and Wang]{zhang2024bidirectional}
Canyu Zhang, Youbao Tang, Ning Zhang, Ruei-Sung Lin, Mei Han, Jing Xiao, and Song Wang.
\newblock Bidirectional autoregessive diffusion model for dance generation.
\newblock In \emph{Proceedings of the IEEE/CVF Conference on Computer Vision and Pattern Recognition}, pages 687--696, 2024{\natexlab{a}}.

\bibitem[Zhang et~al.(2023)Zhang, Rao, and Agrawala]{zhang2023adding}
Lvmin Zhang, Anyi Rao, and Maneesh Agrawala.
\newblock Adding conditional control to text-to-image diffusion models.
\newblock In \emph{Proceedings of the IEEE/CVF international conference on computer vision}, pages 3836--3847, 2023.

\bibitem[Zhang et~al.(2022)Zhang, Cai, Pan, Hong, Guo, Yang, and Liu]{zhang2022motiondiffuse}
Mingyuan Zhang, Zhongang Cai, Liang Pan, Fangzhou Hong, Xinying Guo, Lei Yang, and Ziwei Liu.
\newblock Motiondiffuse: Text-driven human motion generation with diffusion model.
\newblock \emph{arXiv preprint arXiv:2208.15001}, 2022.

\bibitem[Zhang et~al.(2024{\natexlab{b}})Zhang, Li, Cai, Ren, Yang, and Liu]{zhang2024finemogen}
Mingyuan Zhang, Huirong Li, Zhongang Cai, Jiawei Ren, Lei Yang, and Ziwei Liu.
\newblock Finemogen: Fine-grained spatio-temporal motion generation and editing.
\newblock \emph{Advances in Neural Information Processing Systems}, 36, 2024{\natexlab{b}}.

\bibitem[Zhou et~al.()Zhou, Barnes, Lu, Yang, and Li]{zhou2019continuity}
Yi Zhou, Connelly Barnes, Jingwan Lu, Jimei Yang, and Hao Li.
\newblock On the continuity of rotation representations in neural networks.
\newblock In \emph{Proceedings of the IEEE/CVF conference on computer vision and pattern recognition}, pages 5745--5753.

\bibitem[Zhuang et~al.(2022)Zhuang, Wang, Chai, Wang, Shao, and Xia]{zhuang2022music2dance}
Wenlin Zhuang, Congyi Wang, Jinxiang Chai, Yangang Wang, Ming Shao, and Siyu Xia.
\newblock Music2dance: Dancenet for music-driven dance generation.
\newblock \emph{ACM Transactions on Multimedia Computing, Communications, and Applications (TOMM)}, 18\penalty0 (2):\penalty0 1--21, 2022.

\end{thebibliography}
}
\end{document}